\documentclass[aps,prl,twocolumn,showpacs,superscriptaddress,groupedaddress,amsfonts,amsmath,floatfix]{revtex4}

\usepackage{graphicx}
\usepackage{color}
\usepackage{subfigure}
\usepackage{bm}
\usepackage{epstopdf}

\begin{document}
\title{Exotic quantum spin models in spin-orbit-coupled Mott insulators}

\author{J. Radi\'{c}}
\affiliation{Joint Quantum Institute and Department of Physics, University of Maryland, College Park, Maryland 20742-4111, USA}

\author{A. Di Ciolo}
\affiliation{Joint Quantum Institute and Department of Physics, University of Maryland, College Park, Maryland 20742-4111, USA}
\affiliation{Department of Physics, Georgetown University, Washington, D.C. 20057, USA} 

\author{K. Sun}
\affiliation{Joint Quantum Institute and Department of Physics, University of Maryland, College Park, Maryland
20742-4111, USA}
\affiliation{Condensed Matter Theory Center, Physics Department, University of Maryland, College Park, Maryland 20742-4111, USA}

\author{V. Galitski}
\affiliation{Joint Quantum Institute and Department of Physics, University of Maryland, College Park, Maryland
20742-4111, USA}
\affiliation{Condensed Matter Theory Center, Physics Department, University of Maryland, College Park, Maryland 20742-4111, USA}

\date{\today}

\begin{abstract}

We study cold atoms in an optical lattice with synthetic spin-orbit coupling
in the Mott-insulator regime. We calculate the parameters of the corresponding
tight-binding model using Peierls substitution and ``localized Wannier states
method'' and derive the low-energy spin Hamiltonian for fermions and bosons. 
The spin Hamiltonian is a combination of Heisenberg model, quantum compass model and
Dzyaloshinskii-Moriya interactions and it has a rich classical phase
diagram with collinear, spiral and vortex phases.

\end{abstract}


\maketitle

Since the first experimental realization of Bose-Einstein condensate (BEC),
cold atoms have proven to be an excellent playground for studying many-body
physics~\cite{Lewenstein2007,Bloch2008} and many interesting phenomena take
place for strongly interacting atoms in an optical lattice. These studies started with an experimental observation of superfluid to Mott-insulator phase transition~\cite{Greiner2002} which was followed by experimental and theoretical work on both Bose and Fermi gases in 
lattices of different dimensionality and in various parameter regimes~\cite{Lewenstein2007,Bloch2008}. 
The key features of cold atoms in optical lattices are the excellent tunability of parameters and the fact that the sample is almost perfectly described by the Hubbard model in the deep lattice regime~\cite{Bloch2008}.
  Since it is well known that the Hubbard model is mapped to an effective spin Hamiltonian for Mott-insulator phases with integer filling~\cite{MacDonald1988}, it is clear that cold atoms can be used to 
``engineer'' various quantum spin systems~\cite{Svistunov2003,Duan2003} from those described by the Heisenberg model to more exotic ones, 
like the Kitaev model~\cite{Kitaev2006}. In designing effective spin systems different tools like polar molecules~\cite{Zoller2006} and tilted optical lattices~\cite{Greiner2011} can also be used.
In recent years there has been a lot of interest in creating artificial Abelian and non-Abelian gauge fields in cold-atom systems~\cite{Dalibard2011} and successful experimental realizations of synthetic magnetic~\cite{Lin2009,Bloch2011} and electric field~\cite{Lin_electric} and spin-orbit coupling (SOC)~\cite{Lin2011} have been reported. 
  The role of SOC in cold atoms has been extensively studied following
  the theoretical proposal for the creation of artificial  
  SOC~\cite{Stanescu2007,Ruseckas2005} and rich phase diagrams have been found in BECs~\cite{Stanescu2008, Wang2010,Santos2011} and fermionic systems~\cite{Sato2009,Iskin2011,Takei2012}.

 In this letter we combine optical lattice and SOC which, in the deep lattice
 regime, leads to tight-binding description with non-zero ``spin-flip''
 hopping between neighboring sites~\cite{Goldman2009}. We show that in the Mott-insulator phase with integer filling the system is described by an interesting effective spin Hamiltonian which is a combination 
of Heisenberg model, quantum compass model and Dzyaloshinskii-Moriya terms.
  We note that combination of an optical lattice and SOC has already been considered with a purpose of studying superfluid-insulator transition~\cite{Grass2011}, topological phase transitions~\cite{Goldman2012} and BEC dynamics~\cite{Larson2010}.
In the context of solid-state physics spin models resulting from Mott-insulators with strong SOC were studied in Ref.~\cite{Shekhtman1992,Shekhtman1993,Zheludev1999,Jackeli2009}.
 
  We study a two-dimensional system of pseudospin-$1/2$ atoms on a square
optical lattice with synthetic SOC. The single-particle physics is described by the Hamiltonian:
\begin{equation}
\begin{split}
 & H_0=\left[ \frac{{\mathbf p}^2}{2m} + V_x \sin^2(Kx) + V_y \sin^2(Ky)
    \right] \check{{\mathbf 1}} \\  
   &\quad +\alpha \check{\sigma}_x p_x + \beta \check{\sigma}_y p_y,
\end{split}
\label{the_model}
\end{equation}
with ${\mathbf p}$ and $m$ being the atomic momentum and mass; $V_x$, $V_y$ the lattice depth in
$x$ and $y$ direction, $K=\pi/a$ ($a$ is the 
lattice spacing), $\check{{\mathbf 1}}$ the $2 \times 2$ unit matrix, $\check{\sigma}_i$ Pauli matrices; $\alpha$ and $\beta$ characterize the SOC.
Since (\ref{the_model}) is invariant under lattice translations, its
eigenstates have Bloch-wave form: 
$\vec{\psi}_{{\mathbf  k},n}({\mathbf r})=\exp(i {\mathbf k} \cdot {\mathbf r}) \vec{u}_{{\mathbf k},n}({\mathbf r})$ and $\vec{u}_{{\mathbf k},n}({\mathbf r}+{\mathbf R})=\vec{u}_{{\mathbf k},n}({\mathbf r})$, 
${\mathbf R}$ being a lattice vector. 
	We are interested in the deep lattice regime in which pairs of energy bands are well separated and the 
low-energy physics is captured by the lowest pair of bands which touch at ${\mathbf k}=(0,0)$, $(0,K)$, $(K,0)$ and $(K,K)$ in the energy spectrum. 
In this regime the system is well described by the tight-binding approximation in which the Hilbert space 
is spanned by states localized 
on individual lattice sites and the tunneling exists only between nearest-neighbor sites. 
There are two localized states per site ($\rvert W_{\mathbf R}^{1} \rangle$, $\rvert W_{\mathbf R}^{2} \rangle$), hence 
we have two effective particle species.
This is the most general tight-binding description of (\ref{the_model}):
\begin{equation}
  H_{\rm T}= - \sum_{{\mathbf r} i j} \sum_{\gamma=x,y}\big[ a_{{\mathbf r}, i}^{\dagger} 
 T_{\gamma}^{(i,j)}  a_{{\mathbf r}+\bm{\eta}_\gamma, j} + \rm H.c. \big],
\label{hopping_hamiltonian}
\end{equation}
where $a_{{\mathbf r}, i}^{\dagger}$ ($a_{{\mathbf r}, i}$) creates (annihilates) a particle in the state $\lvert W_{\mathbf r}^{i} \rangle$,  $T_{\gamma}^{(i,j)}= -\langle W_{\mathbf r}^{i} \lvert H_0 \rvert W_{{\mathbf r}+\bm{\eta}_{\gamma}}^{j}\rangle$ are the tunneling matrices and $\bm{\eta}_\gamma=a \hat{\gamma}$.  
In finding the elements of $T_\gamma$ corresponding to $H_0$,
we use Peierls substitution and ``localized Wannier states (LWS) method''. 
We write SOC in a gauge-field form: ${\mathbf p}^2/(2m)+\alpha \check{\sigma}_x p_x + \beta \check{\sigma}_y p_y=(\mathbf p - \mathbf A)^2/(2m) + const.$ with $\mathbf A=(-m \alpha \check{\sigma}_x, 
-m \beta \check{\sigma}_y)$ and may use Peierls substitution to find tunneling matrices
\begin{equation}
 T_{\gamma}= t_{\gamma} e^{-i a A_{\gamma}} = t_{\gamma} e^{i \theta_{\gamma} \check{\sigma}_{\gamma}},
 \quad \gamma=x,y 
\label{hopping_Peierls}
\end{equation}
where $t_\gamma$ are tunneling coefficients in the $\gamma$-direction in the absence of SOC, $\theta_x=\pi \alpha^{\prime}/2$, $\theta_y=\pi \beta^{\prime}/2$; $\alpha^{\prime}$ and $\beta^{\prime}$ are dimensionless
SOC strengths: $\alpha^{\prime}= 2 m \alpha /(\hbar K)$, 
$\beta^{\prime}=2 m \beta /(\hbar K)$.  
However, Peierls substitution is only an approximation, valid for SOC weak
with respect to the kinetic plus lattice part of $H_0$. This is the case when $\alpha^{\prime} \ll 2 \sqrt{V_x/E_R}$, $\beta^{\prime} \ll 2 \sqrt{V_y/E_R}$,
where $E_R=\hbar^2 K^2/(2m)$ is the lattice recoil energy.
  While the SOC is quite weak in solid-state systems, in cold-atom systems it
is typically very strong and in that case Peierls substitution is not completely valid. For example, if we combine optical lattice with spacing $a=410\ \rm nm$ and Rashba SOC generated by a scheme described in Ref.~\cite{Campbell2011}, we obtain $\alpha^{\prime} \approx 1$, while the usual experimental values of
lattice depth are $\sqrt{V_{x,y}/E_R} \sim 2 - 5$ (for smaller values of $V_{x,y}$ the tight-binding approximation
is not valid). For the SOC scheme experimentally realized~\cite{Lin2011} $\alpha^{\prime} \approx 1.4$.
  Since the validity condition for Peierls substitution is not completely satisfied, we calculate tunneling matrices using LWS method which is more involved and requires numerical approach 
but it does not contain any approximation.
In a system with $N_x N_y$ sites and periodic boundary conditions, Wannier states for a single band are 
defined as~\cite{Wannier1937,Kohn1959}:
$\lvert W_{\mathbf r} \rangle = \frac{1}{\sqrt{N_x N_y}} \sum_{\mathbf k} e^{-i{\mathbf k} \cdot \mathbf r} 
 e^{i\varphi(\mathbf k)} \lvert \psi_{\mathbf k} \rangle$,
where $\lvert \psi_{\mathbf k} \rangle$ are Bloch states and $\varphi(\mathbf k)$ is an arbitrary phase.
In the absence of SOC it is possible to obtain maximally LWS by varying the 
phase $\varphi(\mathbf k)$ of each Bloch state $\lvert \psi_{\mathbf k} \rangle$ \cite{Kohn1959} and in the deep
lattice regime these maximally LWS are well localized on individual sites.   
In the presence of SOC it is generally not possible to have Wannier states localized on individual sites
if these are constructed from Bloch states of a single band. Therefore we consider generalized Wannier states introduced in Ref.~\cite{Vanderbilt1997}:
 $\lvert {W}_{\mathbf r}^{m} \rangle=\frac{1}{\sqrt{N_x N_y}}  
 \sum_{{\mathbf k},n} e^{-i {\mathbf k} \cdot \mathbf r} M_{mn}({\mathbf k}) 
 \lvert \psi_{{\mathbf k},n} \rangle$,
where $M_{mn}({\mathbf k})$ are $2 \times 2$ unitary matrices which mix Bloch states of the two bands. 
We find maximally LWS by minimizing the functional $\Omega = \sum_{n=1}^{2} \left[ \langle
r^2 \rangle_n - \langle {\mathbf r} \rangle^{2}_{n} \right]$  with respect to
matrix elements of $M_{mn}({\mathbf k})$ ($\langle ... \rangle_n$ is an expectation value associated to $\lvert W_{\mathbf 0}^{n} \rangle$). The minimization is done numerically and 
example of an algorithm is given in Ref.~\cite{Vanderbilt1997}. 
  Numerical results show that the tunneling matrices still have the form given in (\ref{hopping_Peierls}) but now the parameters $t_x$, $t_y$, $\theta_x$ and $\theta_y$ are some more general functions of $V_x$, $V_y$, $\alpha$, $\beta$, $K$ and $m$. It can be shown that the structure of tunneling matrices (\ref{hopping_Peierls}) follows from the symmetries of $H_0$.
  
Peierls substitution has the advantage to give an analytical form for
tunneling matrices, however it
does not give any information about the Wannier states, whereas the LWS method explicitly gives states $\lvert W_{\mathbf r}^{1} \rangle$, $\lvert W_{\mathbf r}^{2} \rangle$ which is important in interpreting the experimental data. 
\begin{figure}
\centerline{
\mbox{\includegraphics[width=0.74\columnwidth]{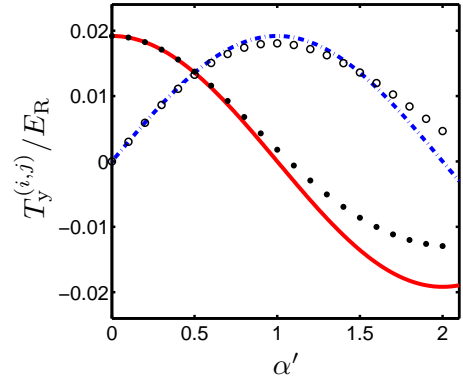}}
}
\caption{(color online)
Tunneling amplitudes $T_{y}^{(1,1)}$ and $T_{y}^{(1,2)}$ obtained by
Peierls substitution (full red and dash-dotted blue line) 
and localized Wannier states method (full dots and empty circles) for different
strengths of Rashba SOC and $V_x=V_y=10\ E_R$.} 
\label{fig_comparison}
\end{figure} 
In Fig.~\ref{fig_comparison} we compare tunneling amplitudes in the
Rashba-coupling case obtained by Peierls substitution and LWS method for
$V_x=V_y=10\ E_R$. They show excellent accord for small $\alpha^{\prime}$ and sizable differences for larger ones.    

Cold atoms in optical lattices are described by the tunneling Hamiltonian (\ref{hopping_hamiltonian}) plus 
interactions~\cite{Bloch2008}:
\begin{equation}
 V=\frac{1}{2} \sum_{{\mathbf r} i j} U_{i j} 
 \colon n_{{\mathbf r},i} n_{{\mathbf r},j} \colon,
\label{interactions}
\end{equation}
where $n_{{\mathbf r},i}$ is a number of particles in state 
$\lvert W_{\mathbf r}^{i} \rangle$, $U_{ij}$ are interaction coefficients and :(...): 
denotes normal ordering of creation and annihilation operators. 


We are interested in the Mott-insulator regime with $t_x/U_{ij} \ll 1$, $t_y/U_{ij} \ll 1$  and integer number $\nu$ of atoms per site ($\nu=1$ for fermions and any
integer $\nu$ for bosons). 
In this case interactions (\ref{interactions}) are the dominant part in the full Hamiltonian 
$H=H_{\rm T} + V$ and we may treat the problem 
perturbatively by taking $V$ as starting Hamiltonian and $H_{\rm T}$ as perturbation. 
The ground state of $V$ is a state with uniform distribution of
atoms 
and the ground state degeneracy is very large since there are two states per site that 
atoms can occupy. The perturbation $H_{\rm T}$ couples the ground-state manifold and
excited states of $V$ and the resulting low-energy effective Hamiltonian
can be calculated in the second order of perturbation theory~\cite{Svistunov2003,Duan2003}:
%
\begin{equation}
 (H_{\rm eff})_{\alpha \beta}= - \sum_{\gamma} \frac{(H_{\rm T})_{\alpha \gamma} (H_{\rm T})_{\gamma \beta}}
 {E_{\gamma} - (E_{\alpha} + E_{\beta})/2},
\label{perturbation_theory}
\end{equation}
where $\alpha$ and $\beta$ label states in the ground-state manifold, while
$\gamma$ labels the excited states of $V$.

 We first calculate the effective low-energy Hamiltonian for fermions.
Since two fermions cannot occupy the same quantum state
the only interesting regime is when $\nu=1$. The only relevant interaction
coefficient is $U_{12}=U$ and the excited states 
of $V$ are those with two fermions of different species at the same site. 
Now it is convenient to introduce isospin operators:
$S_{\mathbf r}^x=(a_{\mathbf r,1}^{\dagger} a_{\mathbf r,2} + a_{\mathbf r,2}^{\dagger} a_{\mathbf r,1})/2$,
$S_{\mathbf r}^y=-i(a_{\mathbf r,1}^{\dagger} a_{\mathbf r,2} - a_{\mathbf r,2}^{\dagger} a_{\mathbf r,1})/2$ 
and $S_{\mathbf r}^z=(n_{\mathbf r,1} - n_{\mathbf r,2})/2$.  
  Using (\ref{perturbation_theory}) we obtain 
\begin{equation}			
\begin{split}
 & H_{\rm F}= \sum_{\mathbf r^{\prime}-{\mathbf r}=\bm{\eta}_x}  \bigg[ J_{\rm h}^{x} 
 {\mathbf S}_{\mathbf r} \cdot 
 \mathbf{S}_{\mathbf r^{\prime}} + J_{\rm cm}^x S_{\mathbf r}^{x} S_{\mathbf r^{\prime}}^{x} 
+ {\mathbf D}^x \cdot \left( {\mathbf S}_{\mathbf r} \times {\mathbf S}_{\mathbf r^{\prime}} 
 \right) \bigg] \\
 &\quad + \sum_{\mathbf r^{\prime} -{\mathbf r}=\bm{\eta}_y}  \bigg[ J_{\rm h}^{y} \mathbf{S}_{\mathbf r} \cdot 
 \mathbf{S}_{\mathbf r^{\prime}} + J_{\rm cm}^y S_{\mathbf r}^{y} S_{\mathbf r^{\prime}}^{y} 
+ {\mathbf D}^y \cdot \left( {\mathbf S}_{\mathbf r} \times {\mathbf S}_{\mathbf r^{\prime}} 
 \right) \bigg], 
\end{split}
\label{H_F_lattice}
\end{equation}
%
where $J_{\rm h}^i= J_i \cos(2 \theta_i)$, $J_{\rm cm}^i= 2 J_i \sin^2 \theta_i$, 
${\mathbf D}^x=J_x \sin(2 \theta_x) \hat{x}$, ${\mathbf D}^y=J_y \sin(2 \theta_y) \hat{y}$, $J_i= 4 t_{i}^{2}/U$
and $\theta_i$ are introduced in (\ref{hopping_Peierls}).
The Hamiltonian is a combination of Heisenberg model,
compass model and Dzyaloshinskii-Moriya-type terms;
for $\theta_x=\theta_y=0$ (no SOC) and $J_x=J_y$ we recover the Heisenberg model~\cite{MacDonald1988,Svistunov2003,Duan2003}. 
\

  Next we find the effective low-energy Hamiltonian for bosons, and now the
  number of atoms per site $\nu$ can be greater than one. The calculation for
  any $\nu$ and general interaction coefficients is very cumbersome, however it simplifies  
for $\delta U_{mn} \ll U$, where $U_{mn}=U+\delta U_{mn}$ or when $\nu=1$ 
\cite{Svistunov2003}. We express the Hamiltonian in terms of
spin-$\nu/2$ operators defined in the same way as in the previous case. For
$\delta U_{mn} \ll U$, the Hamiltonian in the first order of $\delta U/U$ is
$H_{\nu B}=-H_{F \rm} + \sum_{\mathbf r} \big[ A (S_{\mathbf r}^{z})^2 -
  h_{\nu} S_{\mathbf r}^{z} \big]$ where $A=\left(U_{11}+U_{22}-2 U_{12} \right)/2$ and $h_{\nu}= -(\nu-1) \left(U_{11}-U_{22} \right)/2$.
  

For $\nu=1$ and general $U_{mn}$ we obtain
\begin{equation}
\begin{split}		
 & H_{\rm 1B}= - \sum_{i=x,y} \ \sum_{\mathbf r^{\prime}-{\mathbf r}=\bm \eta_i}  \bigg[ J_{\rm h}^{i} 
 {\mathbf S}_{\mathbf r} \cdot 
 \mathbf{S}_{\mathbf r^{\prime}} + J_{\rm cm}^i S_{\mathbf r}^{i} S_{\mathbf r^{\prime}}^{i}
+ J_{z}^{i} S_{\mathbf r}^{z} S_{\mathbf r^{\prime}}^{z} \\
 & \qquad \quad + {\mathbf D}_{B}^{i} \cdot \left( {\mathbf S}_{\mathbf r} \times {\mathbf S}_{\mathbf r^{\prime}} 
\right) \bigg] - h_1 \sum_{\mathbf r} S_{\mathbf r}^z
\end{split}
\label{H_1B_lattice}
\end{equation}
where $J_{z}^i= (u_1+u_2-2) J_{\rm h}^i$, $J_{\rm cm}^i= 2 J_i \sin^2 \theta_i$, ${\mathbf D}_{B}^{i}=(u_1+u_2) {\mathbf D}^i/2$,
$h_1=(u_1-u_2) \left( J_x \cos^2 \theta_x + J_y \cos^2 \theta_y \right)$,
$u_1=U_{12}/U_{11}$, $u_2=U_{12}/U_{22}$ and $J_x$, $J_y$ are given in (\ref{H_F_lattice}) with $U$
replaced by $U_{12}$. 
Since the atoms usually used in experiments have almost spin-independent interactions
we assume $U_{ij} = U$ ($u_1 = u_2 = 1$): this simplifies the Hamiltonian and yields $H_{1 \rm B}= - H_{\rm F}$. 
\begin{figure}
\centerline{
\mbox{\includegraphics[width=0.95 \columnwidth]{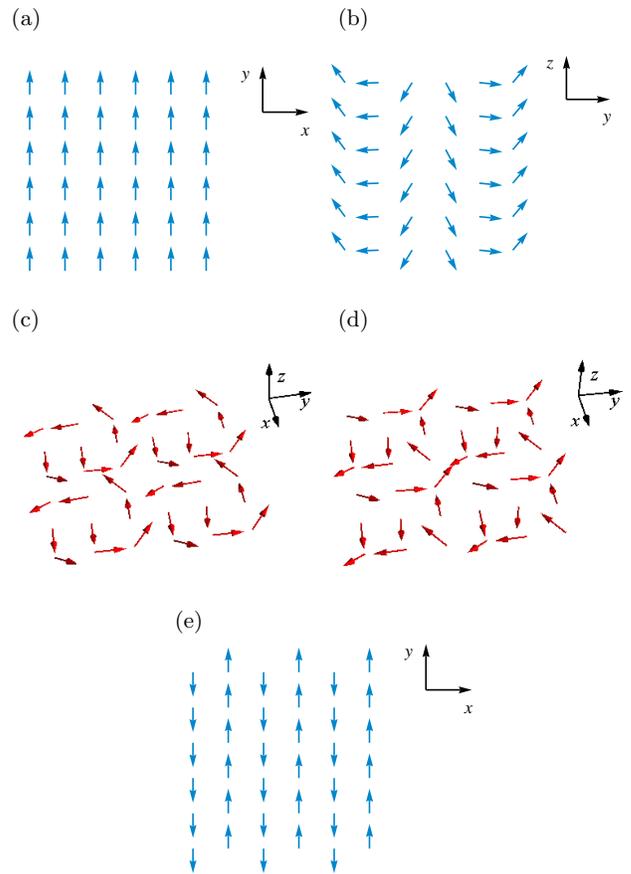}}   
} 
\caption{(color online)
Spin textures:
(a) ferromagnet ($\theta_x=\theta_y=0$); (b) spiral wave ($\theta_x=0.5$, $\theta_y=0.2$); (c) vortex phase ($\theta_x=\theta_y=1$); (d) antivortex phase ($\theta_x=2.14$, $\theta_y=0.96$); (e) stripes ($\theta_x=1.6$, $\theta_y=0.7$). 
Phases (a), (b) and (e) are coplanar and shown in a two-dimensional
representation.}   
\label{phases}
\end{figure}
\begin{figure}
\centerline{
\mbox{\includegraphics[width=0.8 \columnwidth]{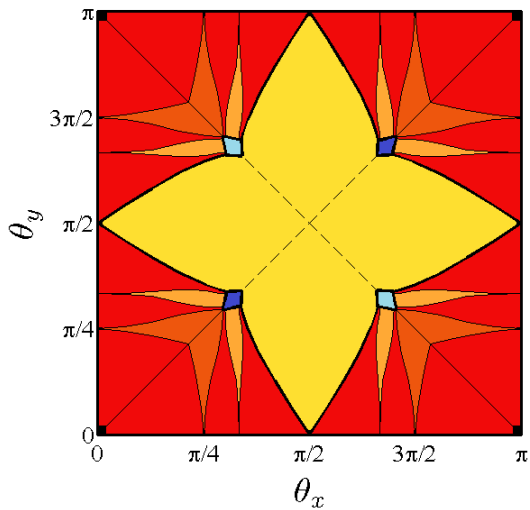}} 
} 
\caption{(color online)
Classical phase diagram of the Hamiltonian $H_{1 \rm B}$ (see text for details): ferromagnet (black corner dots); spiral waves [dark orange (commensurate with four-sites periodicity), light orange (commensurate with three-sites periodicity), red (others)]; stripes (yellow); vortex phase (dark blue) and antivortex phase (light blue).} 
\label{phase_diagram}
\end{figure} 

We intend to find the classical zero-temperature phase diagram of (\ref{H_1B_lattice}) with $u_1=u_2=1$.    
Some previous papers presented models combining Heisenberg,
Dzyaloshinskii-Moriya and compass-model interactions~\cite{Shekhtman1992,
  Shekhtman1993, Zheludev1999} but they did not provide a complete phase diagram, neither at a classical level, usually considering only small SOC. 
In our approach we treat the spins ${\mathbf S}_i$ as classical vectors
and we aim to find the configurations $\lbrace {\mathbf S}_i
\rbrace$ which minimize the energy, with constant $J_x=J_y$. 
We did our computations usually on $60\times 60$-site lattices and finite-size effects are negligible. 
In Fig.~\ref{phases} we show the phases and in Fig.~\ref{phase_diagram} the corresponding phase diagram. 
We obtain two Ising-type phases [ferromagnet (Fig.~\ref{phases}a) and stripes (Fig.~\ref{phases}e)], 
coplanar spirals (Fig.~\ref{phases}b) 
and three-dimensional ordered phases with vortices (Fig.~\ref{phases}c) or
antivortices (Fig.~\ref{phases}d). 
In describing our results, it is helpful to focus on a so called
``basic region'' given by the triangle $\theta_y \le \theta_x \le \pi/2$: the solutions for other parameters can be obtained by simple mappings,
e.g. ground-state configurations in $\theta_x \le \theta_y \le \pi/2$ region
are obtained by simultaneous $\pi/2$-rotation of spins and sites of ground states in the ``basic region''. 
Upon activating SOC, the ferromagnet is immediately replaced by spiral waves,
whose spatial periodicity reduces from several to three sites upon increasing
$\theta_x$ and $\theta_y$; we found both commensurate and incommensurate
waves.  When the compass-model term becomes dominant over the
Dzyaloshinskii-Moriya one, another coplanar phase appears, the ferromagnetic
stripe order, either directly or via the three-dimensional ordered phases. 
We always find non degenerate classical ground states, except along the dashed lines (Fig.~\ref{phase_diagram}) which indicate points in the parameter space with a continuous degeneracy of classical ground states. 
However, we expect this degeneracy to be removed by slight deviations of the
realistic engineered SOC with respect to the Rashba-Dresselhaus form of the
coupling~\cite{Campbell2011}.
The dashed lines also represent the boundaries between phases with stripes of different orientation, {\em i.e.} between the phase shown in
Fig.~\ref{phases}e and the one obtained by rotating the sites and spins of the latter by $\pi/2$ around the $z$-axis. 
The vortex phase (Fig.~\ref{phases}c) takes place along the diagonal
$\theta_x=\theta_y$: vortices are left-handed in the region with smaller SOC and right-handed in the one with larger SOC. 
The antivortex phase (Fig.~\ref{phases}d) is found along the diagonal $\theta_y=\pi-\theta_x$ and the configuration (d) is obtained from the phase (c) by a transformation which reflects sites (but not spins) with respect to the $x$-axis. 
For a better identification of the phase properties, we consider their behavior with respect to the breaking of the translational symmetry of (\ref{H_1B_lattice}). 
While all the phases (except the ferromagnet) break this symmetry, they do it in a different way, {\em i.e.} the stripe phase in Fig.\ref{phases}e
is not invariant under one-lattice-site translation along $x$-direction, but it is invariant under two-lattice-sites translations in $x$-direction and under one-lattice-site translation in $y$-direction; 
the phases with vortices or antivortices are not invariant under one- and two-lattice-sites translations in $x$ and $y$-direction, but they are invariant under three-lattice-sites translations.
Then we can understand the evolution from stripe to vortex phase as a transition in which two-lattice-sites translational symmetry becomes broken. The same reasoning applies for the rest of the phase diagram.    

It is important to emphasize that the classical analysis yields no gapless modes in the whole parameter region except for the diagonal lines ($\theta_y=\theta_x$, $\theta_y=\pi-\theta_x$) in the stripe region. The absence of these gapless modes provides stronger guidelines for further analysis in a semiclassical or quantum approach.     

In summary, we studied the effects of spin-orbit coupling in cold atoms in an
optical lattice in the Mott-insulating regime. We derived the tight-binding
model using Peierls substitution 
and Localized Wannier State method 
and obtained the effective low-energy Hamiltonian for fermions and bosons:
this takes the form of an exotic spin model with Heisenberg, compass-model and Dzyaloshinskii-Moriya interactions. We determined the classical phase diagram for this model and showed that the interplay between the different interactions is responsible for a large variety of phases: ferromagnet, spirals, stripes, three-dimensional vortex and antivortex phases.
We expect that our classification of ground states could generally survive in a quantum approach; in fact, except for some particular cases we mentioned in the discussion, 
on the classical level there are no degeneracies which would be lifted by quantum fluctuations. 

  We thank W.S. Cole, S. Zhang, A. Paramekanti and N. Trivedi
for discussions. J.R. thanks Stephen Powell and Qinqin Lu for discussions.
This research was supported by US-ARO, JQI (A.D.C. and V.G.), ARO-MURI (J.R.) 
and JQI-NSF-PFC (K.S.). 


\bibliography{spinH_references}

\end{document}